\documentclass[%
preprint%
,secnumarabic%
,amssymb, amsmath,nobibnotes, aps,showpacs]{revtex4}
\usepackage{epsfig}%
\usepackage{graphicx}%
\expandafter\ifx\csname package@font\endcsname\relax\else
 \expandafter\expandafter
 \expandafter\usepackage
 \expandafter\expandafter
 \expandafter{\csname package@font\endcsname}%
\fi

\begin{document}

\title{Leptogenesis  by curvature coupling of heavy neutrinos }%

\author{Gaetano Lambiase$^{a,b}$ and Subhendra Mohanty$^c$}%
\affiliation{$^a$ University of Salerno,  Italy.\\
$^b$INFN , Sezione di Napoli, Italy.}
\affiliation{$^c$Physical Research Laboratory, Ahmedabad 380009, India.}
\def\be{\begin{equation}}
\def\ee{\end{equation}}
\def\al{\alpha}
\def\bea{\begin{eqnarray}}
\def\eea{\end{eqnarray}}

\begin{abstract}

We introduce a CP violating coupling between the Ricci curvature and the heavy right handed neutrinos. This splits the Majorana masses of the left and the right handed heavy neutrinos. When the heavy neutrinos decay, their decay rates are different into left and right chirality light neutrinos. A time varying non-zero Ricci curvature can give rise to a net lepton asymmetry. The source of a non-zero curvature in a radiation era is the quantum fluctations of primordial fields.

\end{abstract}


\maketitle
\section{Introduction}

The origin of the baryon number asymmetry in the Universe is, as well known, a still open problem of
the particle physics and cosmology \cite{kolb}.
The successful prediction of the light element abundances by big bang nucleosynthesis
\cite{Copi,burles} and the observations of Cosmic Microwave Background anisotropies combined with the large structure of the
Universe \cite{wmap,bennet}
show that the baryon to photon number of the universe, i.e. the parameter that characterize such a a asymmetry,  is of the order
 \begin{equation}\label{etaexp}
\eta \equiv \frac{n_B-n_{\bar B}}{s}\lesssim (
9.2 \pm 0.5)\,\, 10^{-11}\,,
 \end{equation}
where $n_B$ ($n_{\bar B}$) is the baryon (antibaryon) number
density, and $s$ the entropy of the Universe ($s=2\pi^2g_{*s}T^3/45$, $g_{*s}$ counts the total
degrees of freedom for particles that contribute to the entropy of
the Universe, and finally $g_{*s}$ takes values very close to the total
degrees of freedom of effective massless particles $g_*$).
Explaining $\eta_B$ in terms of known parameters is one of the major problems in cosmology.
In a (CPT invarinat) theory, the baryon asymmetry can be explained provided that the Sakharov' conditions are satisfied \cite{sakharov}:
1) there must exist processes that violate the baryon number; 2) the discrete symmetries
C and CP must be violated; 3) departure from thermal equilibrium.
However, the Sakharov's conditions may be relaxed in some circumstances \cite{dolgov}.
As shown in \cite{cohen}, in fact, a dynamical violation of CPT (which implies a different spectrum of particles and antiparticles)
may give rise to the baryon number asymmetry also in a regime of thermal equilibrium (see also \cite{steinhardt}).

In this paper raise the possibility that the baryon/lepton asymmetry in the Universe can be generated by a CP violating curvature coupling of the heavy Majorana neutrinos \cite{lambiaseJCAP}
  \begin{equation}
{\cal L}_{\diagup{\!\!\!\!\!\!C\!\!P} }= \sqrt{-g}\,\beta \, R
\bar \psi \,i\gamma_5 \psi\,, \label{cpv0}
\end{equation}
where a non-zero Ricci curvature is generated in the radiation era by back-reaction effects of quantum fields.
The dimensional $\beta$ will be assumed to be of order $M_P^{-1}$, where $M_P=G^{-1/2}$ is the Planck mass.
Quantum effects, in particular at the very early time, cannot be ignored because they may modify the dynamics of the Universe evolution. General relativity, therefore, requires some modifications
in order to account for them. In absence of a complete quantum field theory of gravitation, one works assuming a semiclassical theory of gravity.
In this context, the Einstein field equations read \cite{birrell,anderson}
 \begin{equation}\label{Einsteineqs}
    R_{\mu\nu}-\frac{1}{2}g_{\mu\nu}R=\frac{8\pi}{M_P^2}\left(T^{(cl)}_{\mu\nu}+\langle T^{(QM)}_{\mu\nu}\rangle\right)
 \end{equation}
where $T^{(cl)}_{\mu\nu}$ is the stress energy-momentum tensor for the classical field, $T_{\mu\nu}^{(QM)}$ represents the energy momentum tensor operator for quantum fields, and finally $\langle T_{\mu\nu}^{(QM)}\rangle=\langle 0 | T_{\mu\nu}^{(QM)}|0 \rangle$ represents the
regularized expectation value of $T_{\mu\nu}^{(QM)}$. During the radiation dominated era, the trace of the classical energy momentum tensor
vanishes, $T^{(cl)}=0$. However, owing to the presence of the quantum corrections $\langle 0 | T_{\mu\nu}^{(QM)}|0 \rangle$,
it follows that the trace is nonvanishing. This anomaly
comes from the infinite counterterms that must be add to the gravitational action to make the trace finite, and it
is responsible, as we will see, of the generation of the gravitational baryon asymmetry.

The lepton-asymmetry gets frozen-in when the GUT scale lepton-number violating processes decouple.
Baryon asymmetry can then be generated from this lepton-asymmetry \cite{fukugita,luty,Flanz} by the electro-weak sphaleron processes \cite{rubakov}. Sphaleron
processes conserve $(B-L)$ so a lepton asymmetry generated in the GUT era can be converted to baryon asymmetry of the same magnitude
\cite{fukugita}.

\section{CP odd gravitational coupling of Majorana neutrinos}

We consider the $CP$ violating interaction
between fermions and the Ricci curvature described by the dimension-five
operator (\ref{cpv0}) \cite{lambiaseJCAP}.
This operator is invariant under Local Lorentz transformation and
is even under $C$ and odd under $P$ and conserves $CPT$.
In a non-zero background $R$, there
is an effective $CPT$ violation for the fermions.
The Dirac equation
\begin{equation}
i \gamma^\mu \partial_\mu \psi-M \psi -i \beta R \,\gamma_5 \psi =0\,,
\end{equation}
leads to the dispersion relation
\begin{equation}
E^2 \psi=({\bf p}^2 +M^2 + \beta^2 R^2)\psi -\beta(\gamma_5
\gamma^\mu \partial_\mu R)\psi\,.
\end{equation}
For a spatially homogenous background, the energy levels of the
left and right handed fermions split as
\begin{equation}
E^2 \psi_{\pm}= ({\bf p}^2 +M^2 + \beta^2 R^2  \pm \beta \dot R
)\psi_\pm\,,
\end{equation}
where $\psi_{\pm}=(1/2)(1\pm\gamma_5)\psi$ and over-dot represents
time derivative. Now consider a Majorana neutrino in the chiral
representation $N_M=(N_R, i\sigma_2
N_R^*)^T=(N_R,{N_R}^c)^T$. $\dot R \neq 0$ implies a
spontaneous CPT violation. The energy levels of the
right chirality neutrinos $N_R=\psi_+$ and the left chirality
antineutrinos $(\nu_L)^c=\psi_{-}$ are
 \begin{equation}
 E_{\pm}=\sqrt{{\bf p}^2 +M^2 + \beta^2 R^2}\,  \pm \, \frac{\beta \dot R}{ 2 \sqrt{{\bf p}^2
 +M^2 + \beta^2 R^2}}\,.
 \end{equation}
The effective masses $M_{\pm}\equiv E_\pm({\bf p=0})$ of the left and the right chirality states are different,
 \begin{equation}
 M_{\pm} \simeq M + \frac{\beta^2 R^2 }{2 M}  \pm \frac{\beta}{2M}\, \dot R
 \end{equation}
This mass splitting  between the heavy neutrinos and antineutrinos results in the decays.
The heavy right-handed
Majorana neutrino interactions with the light neutrinos and Higgs  relevant for leptogenesis, are
described by the lagrangian
 \begin{equation}
 {\cal{L}}=- h_{\alpha \beta}(\tilde{\phi^\dagger}~
 \overline{N_{R \alpha}} l_{L \beta})
-\frac{1}{2}N_R^c \, M\,N_R +h.c. \,,
\label{LN}
 \end{equation}
where $ M$ is the right handed neutrino mass-matrix, $l_{L \alpha}=
(\nu_\alpha , e^-_\alpha)_L^T $ is the left-handed lepton doublet
($\alpha $ denotes the generation), $\phi=(\phi^+,\phi^0)^T $ is
the Higgs doublet. In the scenario of leptogenesis introduced by
Fukugita and Yanagida, lepton number violation is
achieved by the decays $N_R \rightarrow \phi + l_L$ and also
${N_R}^c \rightarrow \phi^\dagger + {l_L}^c$. The difference in
the production rate of $l_L$ compared to $l_L^c$, which is
necessary for leptogenesis, is achieved via the $CP$ violation. In
the standard scenario, $n(N_R)=n(N_R^c)$ as demanded by $CPT$, but
$\Gamma(N_R \rightarrow l_L + \phi) \not = \Gamma(N_R^c
\rightarrow l_L^c + \phi^\dagger)$ due to the complex phases of
the Yukawa coupling matrix $h_{\alpha \beta}$, and a net lepton
number arises from the interference terms of the tree-level and
one loop diagrams \cite{luty, Flanz}.

In our leptogenesis scenario there is a difference between the heavy right and left chirality neutrinos at thermal equilibrium due to the CP violating gravitational interaction (\ref{cpv0}).
Now even in the absence of CP violating phases in the couplings, there will be a net lepton asymmetry generated from the decay of unequal number of left and right chirality heavy neutrinos,
The decay rate of a heavy neutrinos is
\begin{equation}
\Gamma_{\pm} = \frac{1}{8 \pi} h^2 M_{\pm}
\end{equation}
The lepton asymmetry will be given by
\begin{equation}
\eta= \frac{\Gamma_+ -\Gamma_-}{\Gamma_+ +  \Gamma_-}= \frac{M_+ - M_-}{M}= \frac{\beta \dot R}{M^2}
\label{eta1}
\end{equation}
The lepton asymmetry will freeze-in at the temperature where the GUT scale lepton number violating processes decouple.
In the next section we show that a time varying Ricci curvature can be generated by quantum fluctuations of the fields in the radiation era.

\section{Time varying Ricci curvature from quantum fluctuations}

The dynamical evolution of the gravitational background is assumed to be described by the FRW metric with curvature $K=0$, i.e. $ds^2=dt^2-a^2(t)(dx^2+dy^2+dz^2)$.
The form of the scale factor $a(t)$ is obtained from the semiclassical Einstein equations (\ref{Einsteineqs}).

The regularized components of the energy-momentum tensor we concern have the form \cite{birrell,opher}
 \begin{equation}\label{Tmunu}
    \langle T_{\mu\nu}^{(QM)} \rangle= k_1\,\,{}^{(1)}H_{\mu\nu}+k_3\,\, {}^{(3)}H_{\mu\nu}\,,
 \end{equation}
where
 \begin{eqnarray}
   {}^{(1)}H_{\mu\nu} &=& 2 R_{;\mu;\nu}- 2g_{\mu\nu}  \Box R +2 R R_{\mu\nu}-\frac{R^2}{2} g_{\mu\nu}\,, \label{H}\\
   {}^{(3)}H_{\mu\nu} &=&  R_{\mu}^{\,\,\alpha}R_{\nu \alpha}-\frac{2}{3}R R_{\mu\nu}-\frac{1}{2}R^{\alpha\beta}R_{\alpha\beta}g_{\mu\nu}+\frac{R^2}{4}g_{\mu\nu}\,,
   \nonumber
 \end{eqnarray}
$\Box=\nabla_\mu \nabla^\mu$, and $;$ stands for covariant derivative.
The coefficients $k_{1, 3}$ are constants. Notice that these coefficients come from the regularization process and their values strictly depend not only on number and types of fields
present in the Universe, but also on the method of regularization. Because the methods of regularization affect the the values of $k_{1, 3}$ and more important because of the uncertainty
of what fields were present in the very early Universe, they can be considered as free parameters \cite{anderson,opher}. A comment is in order. The tensor ${}^{(1)}H_{\mu\nu}$ satisfies $\nabla_\mu
{}^{(1)}H^{\mu}_{\nu}=0$ and is derived by varying the local action: ${}^{(1)}H_{\mu\nu}=\displaystyle{2}\sqrt{-g}\frac{\delta}{\delta g_{\mu\nu}}\int d^4 \sqrt{-g} R^2$.  To cancel the infinities in $\langle T^{(QM)}\rangle$
one has to add infinite counterterms in the Lagrangian density describing the gravitational fields. One of these counterterms if of the form $\sqrt{-g}C R^2$, and due to (the logarithmically divergent) constant $C$, the coefficients $k_1$ can take any value, and can be fixed experimentally \cite{k1}. As regards ${}^{(3)}H_{\mu\nu}$, it is covariantly conserved only for conformal flat spacetimes, and cannot be derived by means of the variation of a local action, as for ${}^{(1)}H_{\mu\nu}$. The coefficient $k_3$ takes the form
 \[
 k_3 = \frac{1}{1440\pi^2}\left(N_0+\frac{11}{2}N_{1/2}+31N_1\right)\,.
 \]
For a $SU(5)$ model, for example, the number of quantum fields take the values $N_0=34$, $N_{1/2}=45$, and $N_{1}=24$, so that $k_3 \simeq 0.07$ \cite{opher}.

The explicit expression of the components of ${}^{(1)}H_{\mu\nu}$ and ${}^{(3)}H_{\mu\nu}$ are
 \begin{eqnarray}\label{Hcomponents}
    {}^{(1)}H_{00} & = & 18(2{\ddot H}H+{\dot H}^2+10 {\dot H} H^2)\,, \\
     {}^{(1)}H_{ij} &=& 6\left(2\frac{d^3H}{dt^3}+12{\ddot H} H+14 {\dot H} H^2+7{\dot H}^2\right)g_{ij}\,,
     \nonumber
 \end{eqnarray}
 \begin{equation}\nonumber
    {}^{(3)}H_{00}=3H^4\,, \quad {}^{(3)}H_{00}= H^2(4{\dot H}+3H^2)g_{ij}\,.
 \end{equation}
The process of regularization leads to the a trace anomaly \cite{NoteTrace}
 \begin{equation}\label{Trace}
    \langle T^{(QM)} \rangle = k_3\left(\frac{R^2}{3}-R_{\alpha\beta}R^{\alpha \beta}\right)-6k_1 \Box R\,,
   \end{equation}
that for the FRW Universe reads
  \[
    \langle T^{(QM)}  \rangle = 36 k_1 \left(\frac{d^3H}{dt^3}+7\ddot{H}H+4{\dot{H}}^2+12\dot{H}H^2\right)+
    \]
 \begin{equation}\label{TraceFRW}
 +12k_3H^2\left(\dot{H}+H^2\right)\,.
  \end{equation}
$H={\dot a}/a$ is the Hubble parameter.
The energy-momentum tensor of classical fields is the usual of perfect fluid $T_{\mu\nu}^{(cl)}=$diag $(\rho, -p, -p, -p)$, where $\rho$ is the energy density, $p$ the pressure. They are related by the relation $p=w\rho$, being $w$ the adiabatic index.
During the radiation dominated era, the equation of the state is $p=\rho/3$, i.e. $w=1/3$.
This implies that the trace of the energy-momentum tensor of classical fields vanishes $T^{(cl)}=\rho-3p=0$.
However, the anomaly trace gives a non vanishing contribution.
To evaluate it, we recall that during the radiation dominated era the scale factor evolves as $a(t)=(a_0 t)^{1/2}$, whereas the cosmic time and the temperature are related by the relation $
 \frac{1}{t^{2}}=\frac{32\pi^3 g_*}{90}\frac{T^4}{M_{P}^2}$.
These expressions should be in principle modified by the backreaction effects induced by quantum fields, but as we shall see (and for our purpose) the evolution of the Universe can be described by standard cosmology.

For the FRW, the modified Einstein field equations assume the form
 \begin{equation}\label{EisnteinEqsExplicit1}
    3H^2  =  \frac{8\pi}{M_P^2}\left[\rho+18k_1(2{\ddot H}H+{\dot H}^2+10 {\dot H}H^4)+3k_3 H^4\right]\,,
    \end{equation}
    \begin{equation}\label{EisnteinEqsExplicit2}
    3H^2+2{\dot H} = \frac{8\pi}{M_P^2}\Big[-p+6k_1\Big(2\frac{d^3H}{dt^3}+
    \end{equation}
    \[
    +12{\ddot H}H+14 {\dot H}H^2+7{\dot H}^2\Big) + k_3 H^2(4{\dot H}+3H^2\Big)\Big]\,,
    \]
 from which it follows
 \[
    2H^2+{\dot H}=\frac{8\pi}{M_P^2}\Big[6k_1(\frac{d^3H}{dt^3}+7 {\ddot H}H +4{\dot H}^2+12{\dot H}H^2)
 \]
 \begin{equation}\label{EqH}
    +2k_3 H^2({\dot H}+H^2)\Big]\,.
 \end{equation}
We are looking for solutions of the form
 \begin{equation}\label{H=H0+delta}
    H(t)=H_0(t)+\delta(t)\,,
 \end{equation}
where $\delta(t)\ll 1$ is a perturbation, and $H_0=1/2t$ is the Hubble parameter for a Universe  radiation dominated. During the radiation dominated era, the Ricci curvature vanishes, $R=0$, as well as its covariant and (cosmic) time derivatives ($\Box R=0$, $\nabla_\mu\nabla_\nu R=0$, ${\dot R}=0$, ${\ddot R}=0$). It then follows that ${}^{(1)}H_{\mu\nu}(H=H_0)=0$ (i.e. $\frac{d^3H_0}{dt^3}+7\ddot{H}_0H_0+4{\dot{H}}_0^2+12\dot{H}_0H_0^2=0$) so that it only contains $\delta$-terms and its derivatives. On the contrary, ${}^{(3)}H_{\mu\nu}\neq 0$ when $H=H_0$ owing to the (quadratic) Ricci-terms (in other words we have $H_0^2({\dot H}_0+H_0^2)\sim t^{-4}+F_\delta$, where $F_\delta$ is a function of $\delta$ and its derivatives).
This implies that in the final expression of the baryon asymmetry $\eta_L$, only the $k_3$-terms appear.

Inserting $H$ given in (\ref{H=H0+delta}) into Eq. (\ref{EqH}), to leading order one obtains
 \begin{equation}\label{Eqdelta}
   {\dot \delta}+\frac{2}{t}\delta+ \frac{16\pi k_3}{M_P^2}\frac{1}{t^4} \simeq 0\,,
 \end{equation}
whose solution is
 \begin{equation}\label{Solution}
  \delta(t)\simeq \frac{k_3}{M_P^2}\frac{1}{t^3}-\frac{C}{4M_P^2}\frac{1}{t^4}\,.
 \end{equation}
$C$ is a constant of integration. As it can be seen, the $M_P^{-2}$ suppresses considerably the effects of $\delta$ on the dynamics of the Universe evolution, and these terms wash-out for large $t$.
During the radiation  era $H\simeq1/2t$, it follows that the trace anomaly (\ref{TraceFRW}) reads
 \[
 \langle T^{(QM)} \rangle=-\frac{3k_3}{4t^4}\,.
 \]
From $R=-\displaystyle{\frac{8\pi}{M_P^2}}\langle T^{(QM)} \rangle$ we find that
\bea
\dot R= -\frac{24 \pi k_3}{M_P^2} \frac{1}{t^5}
\eea
the parameter characterizing the heavy neutrino  asymmetry (\ref{eta1}) assumes the form
 \begin{equation}\label{eta1fin1}
 \eta_L(T) = k_3 24 \pi \sqrt{\frac{32^5 \pi^{15} g_*^5}{90^5}} \frac{\beta T^{10}}{M^2 M_P^7}\simeq
 3.6 k_3 \, 10^9 \frac{T^{10}}{M^2}\frac{1}{M_P^8} \,,
 \end{equation}
where $\beta\sim M_P^{-1}$. The heavy neutrino asymmetry freezes at the value $\eta_L(T_D)$ where $T_D$ is the decoupling temperature when the lepton-number violating interactions ($N_R \leftrightarrow N_R^c$) go out of equilibrium. In the following section we will calculate $T_D$ and $\eta_L(T_D)$ in the context of a specific GUT model as an example.

\section{Heavy neutrino asymmetry at thermal equilibrium and Leptogenesis}
In this section we will show how an asymmetry is developed in the heavy neutrinos due to the curvature coupling term. This mechanism works for any model where lepton number violating interactions can take place freely at high temperatures. To be specific we will illustrate our mechanism using a specific GUT model like SO(10). A recent example of leptogenesis due  heavy neutrino decay with CP violation in a SO(10) model is \cite{Abada}.
In standard SO(10) unification, all Standard Model fermions of a given generation together with a right-handed neutrino are in a ${\bf 16}$ representation of SO(10),
\bea
{\bf 16_f} &=& ({\bf 1_f} + {\bf \bar 5_f} + {\bf \bar{ 10}_f})_{SU(5)} \nonumber\\
&=& (N_R + ( L, d^c)+ (Q, u^c, e^c))
\eea
The charged fermion and Dirac neutrino mass matrices receive contributions from Yukawa couplings of the
form $\bf{16_f 16_f H}$ (where ${\bf H = 10_H, 126_H}$ and/or ${\bf 120_H}$).  Majorana masses for the right-handed neutrinos are generated either from
\be
{\bf 16_f\, 16_f \,{\overline {126}}_H} \supset y \,S^\prime\, N_R^c \,N_R
\label{MNr1}
\ee
or from the non-renormalizable operators suppressed by some mass scale $\Lambda$
\be
 \frac{f}{\Lambda}{\bf 16_f\, 16_f \,{\overline{ 16}}_H \,{ \overline{ 16}}_H} \supset \frac{f}{\Lambda}\, S^2\, N_R^c\, N_R\,.
 \label{MNr2}
\ee
When the GUT Higgs fields $S^\prime$ or $S$ acquire a vev , a large Majorana  mass $M$ is generated for $N_R$ which breaks lepton number spontaneously. This following the see-saw mechanism leads to small neutrino masses at low energies. At temperatures larger than the heavy neutrinos and the GUT Higgs masses one there will be helicity flip scattering interactions like $S + N_R \leftrightarrow S+  N_R^c$ which change the lepton number ( as $T> M$ the helicity and the chirality of $N_R$ are same). The interaction rate is
\be
\Gamma(S N_R\leftrightarrow S N_R^c)= \langle n_s \sigma \rangle= \frac{0.12}{\pi} \left(\frac{f}{\Lambda}\right)^2 T^3\,.
\ee
The interactions decouple at a temperature $T_D$ when $\Gamma(T_D)=H(T_D)$ from which we derive the decoupling temperature to be
\be
T_D=13.7 \pi \sqrt{g_*} \left(\frac{\Lambda}{f} \right)^2 \frac{1}{M_P}= 13.7 \pi \sqrt{g_*} \left(\frac{\langle S\rangle^2}{M} \right)^2 \frac{1}{M_P}
\label{TD}
\ee
where we have used $M=f \langle S \rangle/\Lambda$.
Substituting (\ref{TD}) in the expression for lepton asymmetry (\ref{eta1fin1} ) we obtain the value of frozen in lepton asymmetry as
\be
\eta_L(T_D)= k_3 7.8 \times 10^{35} \frac{\langle S \rangle^{40}}{M^{22}M_P^{18}}
\ee
We choose the heavy neutrino mass $M=10^{12}$GeV which is consistent with the atmospheric neutrino scale $m_\nu=0.05$eV with a see-saw relation $m_\nu=m_D^2/M$ with the Dirac neutrino mass scale $ m_D=7$GeV. Taking the GUT Higgs vev $\langle S \rangle =1.2 \times 10^{14} $GeV and $k_3=0.07$ we obtain the frozen-in  asymmetry of the heavy neutrinos as
\be
\eta_L(T_D)= 3.3 \times 10^{-10}
\ee

We note that our scenario for generating heavy neutrino asymmetry is more general than the specific model we have considered above to calculate the temperature of decoupling of the $N_R \leftrightarrow N_R^c$ interactions. We find that if this decoupling temperature is at the GUT scale, $T_d\sim 10^{15}$ GeV then from  (\ref{eta1fin1}) we see that if neutrino mass is $M\sim 10^{12} $ GeV we obtain the required heavy neutrino asymmetry
\be
\eta_L=5.8\times 10^{-10} \left(\frac{T_D}{1.3 \times 10^{15} \,{\rm GeV}}\right)^{10} \left(\frac{ 10^{12}\, {\rm GeV}}{M}\right)^2
\ee

This heavy neutrino asymmetry gets converted into light neutrino asymmetry when they decay, and light neutrino  lepton asymmetry is converted into baryon asymmetry by the action of sphalerons in the electroweak era.

The lepton asymmetry is passed on to the light neutrino sector when the heavy neutrino decays at temperature $T\sim M\sim 10^{12}$GeV. There exists  $\Delta L=2$ interactions, $\nu_L + \phi^0 \leftrightarrow \nu_R + \phi^0$ and
$\nu_R + {\phi^0}^* \leftrightarrow \nu_L + {\phi^0}^*$ that result
from the effective operator ${\cal{L}}_W=\frac{C_{\alpha \beta} }{2M}(\overline{ {l_{L
 \alpha}}^c }~ \tilde{\phi^*})(\tilde{\phi^\dagger}~ l_{L \beta}) +
 h.c.$, where $\alpha, \beta$ denote the generation indices, $l_{L
\alpha}= (\nu_\alpha , e^-_\alpha)_L^T$ is the left-handed lepton
doublet, $\phi=(\phi^+,\phi^0)^T$ is the Higgs doublet (${\tilde
\phi}\equiv i \sigma_2 \phi^*=(-{\phi^0}^*, \phi^-)^T$) \cite{weinberg}. The
The interaction rate for the interaction $\nu_{L
\alpha} +\phi^0 \leftrightarrow \nu_{R \beta} +\phi^0$ is
\be
\Gamma =\langle n_\phi~ \sigma\rangle =
\frac{0.12}{\pi} \frac{|C_{\alpha\beta}|^2 T^3}{M^2}.
\ee
In the electroweak era, when the Higgs field in ${\cal{L}}_W$
acquires a $vev$, $\langle\phi\rangle =(0,v)^T$ (where
$v=174~GeV$), the five dimensional Weinberg operator gives rise to
a neutrino mass matrix $m_{\alpha \beta}= \frac{v^2 C_{\alpha \beta}}{M}$. Hence $C_{\alpha \beta}$ can be written in terms of the heaviest neutrino mass $m_\nu$ as $C\simeq m_\nu M /v^2$. The lepton number violating  interactions decouple
when $\Gamma(T_l)=H(T_l) $ and the decoupling temperature $T_l$ turns out to be
 \begin{equation}\label{decouplingTD}
 T_l= 13.7 \pi~ \sqrt{g_*}~ \frac{v^4}{m_\nu^2 \, M_{P}}\simeq 2\,\, 10^{14}\left(\frac{0.05\text{eV}}{m_\nu}\right)^2\text{GeV}\,,
 \end{equation}

We see that the heavy neutrino decays occur at a temperature $T\simeq M \simeq 10^{12}$ GeV which is much below the temperature $T_l=2 \times 10^{14}$GeV at which the light-neutrino lepton number violating interactions are effective. Therefore the lepton number asymmetry from the decay of asymmetric number of heavy neutrino decays is not washed out by Higgs scattering with light neutrinos.

\section{Conclusions}

In conclusion, we have shown in this paper that a CP violating gravitational interaction between the heavy Majorana neutrinos and the Ricci curvature can generate a lepton asymmetry  in thermal equilibrium in the GUT scale radiation era. The subsequent decays of these heavy neutrinos into the light standard model particles and the conversion of lepton asymmetry into baryon asymmetry can explain the observed baryon asymmetry of the universe.

In order to generate lepton asymmetry we need two ingredients (a) a CP violating interacting term (\ref{cpv0}) of the curvature with heavy Majorana neutrinos and (b) a non-zero value of $\dot R$ in the radiation era to split the energy levels between $N$ and $N^c$. In \cite{lambiaseJCAP} we used the warm inflation scenario to achieve a non-zero $\dot R$ at high temperature (we also studied the the role of beta functions of SU(N) gauge theories in generating a non-zero $\dot R$). In this paper we have shown that a non-zero $\dot R$ can occur in the radiation era from loop  corrections to the trace of the stress tensor and have calculated the contribution of the resulting curvature perturbation to $\dot R$. This mechanism of generating $\dot R$ is more general than the restricted scenarios considered earlier in the earlier study \cite{lambiaseJCAP}.

Some comments are however in order. In principle, one should also taken into account primordial perturbations of the gravitational background (characterized mainly by scalar and tensor perturbations) and of the energy density and pressure, characterized by $\delta \rho=\delta T_0^0$ and $\delta p \delta_i^j=\delta T_i^j$ (see for example \cite{ma}). These perturbations are related as $\delta p = c_s^2 \delta \rho$, where $c_s^2=w+\rho dw/d\rho$ is the adiabatic sound speed squared. For relativistic particles  $w=1/3$ and therefore $c_s^2=1/3$.
As a consequence, the trace of the perturbed energy-momentum tensor vanishes (this is not true
in presence of anisotropic shear perturbations), so that according to the gravitational leptogenesis mechanism, no net baryon asymmetry can be generated.

Moreover, we also compute the energy density of backreaction of quantum fields and compare it with
the energy density of radiation. From Eqs. (\ref{Tmunu}) and (\ref{H}), and using $H\simeq 1/2t$, it follows
 \begin{eqnarray}\label{rhoQFT}
    \langle \rho\rangle &=& \langle T_{00}^{(QM)} \rangle = 18k_1\left(2{\ddot H} H +{\dot H}^2+10{\dot H} H^2\right) \\
   &&  +3k_3 H^4=\frac{3k_3}{8t^4}\,. \nonumber
 \end{eqnarray}
The $k_1$-term vanishes identically. Expressing the total energy density in terms of the scale factor $a(t)=(a_0 t)^{1/2}$ one gets
 \begin{equation}\label{energytot}
    \rho= \rho_r + \langle \rho \rangle = \frac{\rho_0}{a^4}+\frac{A}{a^8}\,,\qquad A\equiv \frac{3k_3}{8}a_0^4\,.
 \end{equation}
where $\rho_r=T_{00}^{(cl)}=\displaystyle{\frac{\pi^2 g_*}{30}T^4}$ is the energy density of the classical radiation.
The ratio between the energy densities $\langle \rho \rangle$ and $\rho_r$ reads
 \begin{equation}\label{ratio}
    r\equiv \frac{\langle \rho \rangle}{\rho_r}=\left(\frac{T}{T_*}\right)^4\,,
 \end{equation}
where we have definite $T_*$ as
 \[
 T_*\equiv \left[\frac{80}{k_3 \pi^4 g_* }\left(\frac{15}{16}\right)^2 \right]^{1/4}M_P \simeq \frac{1}{k_3^{1/4}}10^{18}\,GeV\,.
 \]
For temperatures $T< T_*$ we have that $r<1$, i.e. the energy density of quantum fields is subdominant with respect to the energy density of the radiation.
This is in agreement with the approximation before discussed, where we pointed out that the effects of the backreaction wash out during the Universe expansion. In particular, since the decoupling temperature of heavy neutrinos occurs at GUT scales, $T_D \sim 10^{15}$GeV, we infer $r\sim 10^{-8}\ll 1$ and the backreaction is subdominant over the radiation density.

\section{acknowledgements}

G.L. acknowledges the financial support of
MIUR through PRIN (Prot. 2008NR3EBK 002).


\begin{thebibliography}{99}
\bibitem{k1} It is worth nothing that by making use of the dimensional regularization \cite{duff}, for example, one infers
 \[
   k_1=\frac{1}{1440\pi^2}\left(\frac{1}{2}N_0+3N_{1/2}+6N_1\right)\,.
 \]
\bibitem{NoteTrace} The anomaly trace is typically expressed in term of curvature tensors and their covariant derivatives, as well as mass-terms \cite{birrell,dappiaggi}, i.e.
 \[
 \langle T \rangle = \alpha(N_0, N_{1/2}, N_1)\left(R^{\mu\nu}R_{\mu\nu}-\frac{R^2}{3}\right)+c_1(m) R+
 \]
 \[
 +c_2(m)+c_3\Box R-\sum_{i=1}^{N_0} m_i^2\langle \phi_i^2\rangle
 -\sum_{i=1}^{N_{1/2}} m_i\langle {\bar \psi}_i \psi_i\rangle\,,
 \]
where the coefficients $c_{1, 2}(m)$ are combinations of (power) mass fields, $N_{0, 1/2, 1}$ is the number of the quantum matter of boson, fermion and vector fields. The coefficients $c_i$ are subject to a finite renormalization, becoming free parameters of the theory. For our purpose we can neglect the mass-terms since the fields are relativistic, keeping in mind however, that the procedure of renormalization gives rise to purely geometric terms which appear in the final expression of the trace anomaly.
\end{thebibliography}
\end{document}